# Fourier Transform Infrared microspectroscopy-based super-resolution virtual staining of unlabeled tissues by pixel Diffusion Transformer

Yudong Tian[1], Xiangyu Zhao[1], Yuqing Liu[1], Bofei Yang[1], and Chongzhao Wu[1, *]
[1] Center for Biophotonics, Institute of Medical Robotics, School of Biomedical Engineering, Shanghai Jiao Tong University, Shanghai, China
*Corresponding Author

ABSTRACT: Histological staining is a crucial step in clinical pathological analysis. As the most widely used staining technique, hematoxylin and eosin (H&E) staining enables visualization of tissue and cellular morphology, playing an indispensable role in disease diagnosis. However, the staining process is usually time-consuming, and the chemical reagents involved alter the tissue chemistry, limiting its utility for further downstream analysis. Fourier transform infrared (FTIR) microspectroscopy has shown promising results for characterizing the biochemical components of unstained tissues, but its results are less interpretable for pathologists compared to H&E-stained images. Here, we present a diffusion transformer (DiT)-based pixel super-resolution virtual staining approach to transform low-resolution FTIR microspectroscopic images of the unstained tissues into corresponding high-resolution H&E-stained images. Unlike conventional conditional DiT architectures, this method models the transformation from FTIR images to H&E-stained images as a stochastic Brownian bridge process and directly learns the cross-domain translation in pixel space by means of a large-patch Transformer. When applied to FTIR images of unlabeled human lung tissue samples, the proposed method successfully transforms them into high-resolution H&E-stained images, achieving a 4× pixel-level super-resolution. Additionally, by partitioning images into large patches, our method achieves a fourfold improvement in inference speed compared with traditional U-Net-based diffusion models, without compromising the quality of generated images. This super-resolution virtual staining method provides a rapid and effective solution for generating high-resolution, clinically usable H&E-stained images from infrared spectroscopic images, which can significantly facilitate the incorporation of FTIR microspectroscopy into clinical histological scenarios.

## Introduction

Nowadays, histological staining has become a cornerstone technique in biomedical research and clinical pathology, enabling the visualization of tissue architecture and cellular morphology through selective interaction of dyes with biological components [1–3]. As the most widely used staining method in histology and pathology, hematoxylin and eosin (H&E) staining can clearly differentiate between cellular components in tissues, making it the gold standard in the diagnosis of various diseases, especially cancer [4]. However, traditional H&E staining involves labor-intensive and time-consuming tissue processing steps, typically requiring days to weeks from tissue collection to microscopic examination, which carries the risk of delaying disease diagnosis and treatment. Moreover, the tissue damage introduced during the chemical staining procedures is irreversible, prohibiting additional staining or further molecular analysis on the same section [5,6]. Therefore, developing a rapid and label-free method that can generate high-quality pathological results would be invaluable for clinical diagnostics.

Recently, Fourier transform infrared (FTIR) microspectroscopy has shown promising results for non-destructive and label-free pathological assessment of biological tissues [7,8]. By measuring the mid-infrared light absorption of samples, FTIR microspectroscopy can quantitatively detect natural tissue biomarkers such as proteins, lipids, nucleic acids, etc., which are often correlated with disease progression [9,10]. In recent decades, FTIR has been widely used to improve the diagnosis and treatment of numerous diseases, such as lung cancer [11], liver cancer [12], acute pyelonephritis [8], gastric cancer [13], and so on [14–18]. However, images obtained from FTIR modalities often exhibit lower spatial resolution due to wavelength limitations, and have fundamentally different contrast mechanisms compared to conventional H&E-stained images. As a result, their visual appearance is unfamiliar to clinical pathologists, creating a significant barrier to clinical adoption [19]. In routine diagnostic workflows, H&E-stained images are still required as a reference for interpretation, yet obtaining them necessitates additional staining procedures and involves complex registration processes to ensure accurate spatial alignment, which are technically challenging and error-prone. These practical constraints highlight the need for an efficient approach to bridge the gap between FTIR imaging and conventional histopathology, thereby simplifying the diagnostic workflow and advancing research in infrared pathology.

Over the last decades, the rapid advancements of deep learning (DL)-based approaches have revolutionized the medical image analysis [20,21], demonstrating remarkable performance in various tasks including image super-resolution [22–24], pathological diagnosis [25,26], semantic segmentation [27–29], cross-modality transformation [30,31], and virtual staining [32–39]. In particular, DL-based virtual staining (VS) techniques can generate computationally stained results from various label-free images (e.g., autofluorescence images [32,36,38], quantitative phase microscopy images [40], stimulated Raman scattering images [33,41], imaging mass spectrometry [42], among others), avoiding the complex procedures required in conventional staining protocols and the tissue damage caused by chemical dyes. Early studies on virtual staining primarily relied on convolutional neural networks (CNNs) or generative adversarial networks (GANs) to learn a direct mapping between label-free modalities and stained images. While these methods have demonstrated promising results, they often suffer from inherent limitations, such as blurring of fine structural details, instability during training, and limited generalization to unseen tissue types or imaging conditions [42]. More recently, diffusion models [43,44] have emerged as a powerful alternative for image generation tasks, owing to their strong generative capacity, stable optimization, and superior ability to model complex, multimodal data distributions. By iteratively refining samples through a stochastic denoising process, diffusion models have shown superior performance in preserving fine-grained image structures and color consistency compared with CNN- and GAN-based counterparts. Building upon this framework, Brownian Bridge Diffusion Models (BBDMs) [45] further tailor the diffusion process for image-to-image translation tasks by explicitly constraining the stochastic trajectory between a source image and a target domain, leading to improved conditional controllability and faster convergence. BBDM has been reported to outperform traditional cGAN architecture in various virtual staining applications[32,42]. Nevertheless, most existing diffusion-based virtual staining methods are implemented with U-Net–style backbones, which may limit the model's capacity and efficiency in capturing global, long-range dependencies in images, particularly when handling high-resolution data or when an understanding of complex structural contexts is required. To address this issue, Diffusion Transformers (DiTs) [46] have been proposed by introducing vision Transformer (ViT) architectures into diffusion models, leveraging self-attention mechanisms to explicitly model global context and long-range interactions across the image. While DiT models demonstrate superior capability in capturing high-level semantic consistency, pure Transformer-based designs typically require a relatively small patch size input to generate high-fidelity images with fine-grained structures, which dramatically increases the token length and results in prohibitive computational and memory costs when applied directly in pixel space. Consequently, some studies have explored hybrid architectures that combine Transformers and CNNs to integrate global context modeling with robust local feature representations, enabling diffusion models to achieve both semantic consistency and pixel-level accuracy [47,48]. This paradigm provides an effective and flexible framework for high-quality image generation and translation in complex visual domains. More importantly, a recent study has demonstrated that [49], by predicting clean images rather than noise, a pure vision Transformer operating on large image patches can also effectively perform diffusion modeling directly in pixel space, thereby substantially reducing the computational resources required for running the DiT model and improving inference speed.

In this work, by integrating the Brownian bridge process into the diffusion Transformer model,

we presented a diffusion transformer-based pixel super-resolution virtual staining (DiT-SRVS) model that transforms FTIR microspectroscopic images of unstained tissue samples into H&E-stained higher-resolution images without the need for traditional chemical staining process. Fig. 1 shows the workflow for the DiT-SRVS model, in which a hybrid CNN–Transformer architecture is adopted. The overall model consists of three components: (1) a super-resolution header, (2) a pixel diffusion transformer backbone, and (3) a detail refiner. The super-resolution head is a lightweight convolutional neural network incorporating a pixel-shuffle layer, which upsamples the input low-resolution spectroscopic images and reduces their channel dimensionality to match that of the H&E-stained bright-field images. The pixel diffusion Transformer, built upon a Brownian bridge process, further integrates the low-resolution conditional input with noise estimation to reconstruct high-resolution chemically stained images of unlabeled tissue samples. Finally, a detail refiner composed of a lightweight U-Net network is employed to further enhance the fine-grained local details of the images reconstructed by the DiT backbone. Experimental results on unlabeled human lung tissue demonstrate that, compared with conventional cGAN-based virtual staining methods, our approach produces virtual H&E-stained images with superior visual quality and more accurate color distributions, achieving significant improvements in quantitative metrics. Moreover, the large-patch Transformer substantially reduces the input sequence length and offers higher computational efficiency and faster inference speed than standard U-Net-based diffusion models, which can greatly enhance the clinical throughput of this super-resolution virtual staining technique. In summary, our DiT-SRVS approach provides a powerful and reliable tool for directly and rapidly generating high-resolution H&E-stained images of corresponding unstained tissues from their low-resolution FTIR images, which can help to further accelerate the clinical translation of infrared metabolomics. We believe that this technique will offer significant benefits for FTIR-based histopathological research and will facilitate the integration of FTIR microspectroscopic studies into clinical histological practice.

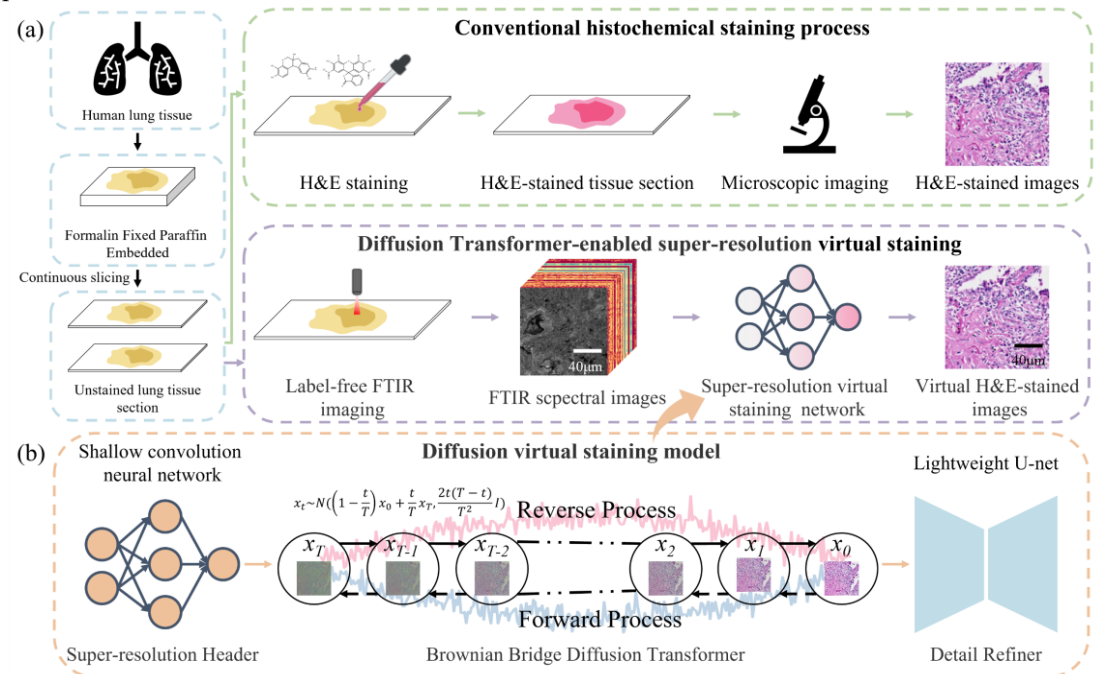

Fig 1. Super-resolution virtual staining of unlabeled tissue sections using diffusion Transformer. (a) Pipelines of traditional hematoxylin and eosin (H&E) histochemical staining and diffusion transformer-based super-resolution virtual tissue staining; (b) Schematic illustration of the workflow of the DiT-SRVS model, which directly learns the transformation between two image domains via a Brownian bridge diffusion process.

## 2. Materials and methods

## 2.1 Sample preparation, histochemical H&E staining and FTIR image acquisition

Lung tissue samples for this research were sourced from formalin-fixed paraffin-embedded (FFPE) tissue blocks acquired by Shanghai Chest Hospital with informed consent from all patients. All the experiment procedures in this study were approved by the Ethics Committee of the Shanghai Chest Hospital [Ethics approval number: KS(Y)22139]. The study involved lung specimens from 6 individual patients. For each patient, two consecutive tissue sections approximately 4 μm thick were sliced from the FFPE tissue blocks. The first section was mounted on a calcium fluoride ($CaF_2$) slide for FTIR imaging. The second section was placed on a glass slide and stained with hematoxylin and eosin (H&E). Both sections were scanned with an OCUS digital pathology scanner to acquire the corresponding bright field images with a spatial resolution of 0.48 μm/pixel (20× objective). Prior to further processing, these bright-field images were downsampled to a resolution of approximately 1.25 μm/pixel.

The FTIR images of lung tissue sections were acquired using a Bruker Hyperion II IR Microscope in the spectral range from 4000 to 900 $cm^{-1}$ at a spectral resolution of 8 $cm^{-1}$. Imaging was performed using a 64*64 pixel focal-plane array detector with a spatial resolution of 5 μm/pixel in transmission mode. Meanwhile, visible-light images of the FTIR imaging regions were captured using the built-in 4x objective lens for subsequent image registration.

## 2.2 Data preparation and pre-processing

### 2.2.1 FTIR image pre-processing

The spectrum in each pixel of the FTIR spectral images was cut into the fingerprint region [50], spanning from 1800 $cm^{−1}$ to 1000 $cm^{−1}$, for subsequent analysis. After wavenumber cutting, the spectrum of each pixel consists of 207 wavenumber channels. Then the FTIR spectra were baseline-corrected using Rubber-band baseline correction with 64 baseline points and band normalized to Amide I (1650 $cm^{-1}$) [51]. Principal component analysis (PCA) was employed for dimension reduction to reduce the training time and memory consumption. And the first five principal components were retained for subsequent model training and evaluation (detailed in the Results section).

### 2.2.2 Multimodal image registration

Due to the substantial differences in spatial resolution and imaging modality, as well as the mismatch in the scan area during image acquisition, the H&E-stained images cannot be directly registered to the FTIR spectral images. To address this challenge, we designed a multi-step registration strategy that leverages the bright-field image of an unstained tissue section for assistance. The workflow consists of the following five steps: (1) The whole-slide image (WSI) of the H&E-stained tissue and the corresponding whole-slide image of the unstained section are registered using the RegWSI tool [52] to achieve global alignment; (2) The visible-light image obtained from the FTIR scan area (as described in Section 2.1) is then used as the template. With the SIFT feature-matching algorithm, we identify the corresponding scan region in the WSI of the unstained section. The matched region is subsequently cropped from the registered H&E WSI to obtain a coarse H&E–FTIR paired image; (3) The FTIR spectral image is processed by integrating the protein-related amide I/II band (1690 - 1490 $cm^{−1}$) to generate a corresponding grayscale image, which is then upsampled by a factor of four using bilinear interpolation to match the spatial size of the H&E image. The H&E image obtained in step (2) is further registered to this grayscale image; (4) Both the H&E image and the grayscale FTIR-derived image are divided into 512×512 patches (The raw FTIR spectral image is similarly partitioned into 128×128 patches.). For each paired image patch, the H&E image patch was aligned to the corresponding FTIR grayscale image patch at the pixel level using an iterative elastic pyramid cross-correlation registration method; (5) For each registered H&E patch (or FTIR spectral patch) obtained in step (4), the central 256×256 (or 64×64) region is cropped and used as the final paired dataset for model training and testing.

**2.2.3 Dataset preparation and division**

The training and testing dataset comprised paired FTIR spectral images and their corresponding bright-field histochemically stained H&E images obtained in section 2.2.2. In this work, we collected a total of 1312 paired FTIR-H&E microscopic image patches (each with 256×256 pixels) patches from six patients. Among them, 1,168 pairs from five patients were used for model training, while the remaining 144 pairs from the last patient were reserved for blind testing. And the paired image FOVs were further randomly cropped to a size of 192×192 pixels during each training epoch. In addition, image pairs that were blurred or exhibited severe misalignment were manually excluded from the dataset to during the training and quantitative evaluation stages.

## 2.3 Brownian bridge diffusion process and Network architecture

As illustrated in Fig. 2, the super-resolution virtual staining framework consists of two networks: a super-resolution header and a pixel diffusion transformer. The super-resolution header is a lightweight convolutional neural network (CNN) composed of two convolutional layers and a pixel-shuffle layer (as depicted in Fig. 2b), which is responsible for transforming the low-resolution FTIR images of unlabeled tissue $y_0 \in R^{\frac{H}{N} \times \frac{W}{N} \times 5}$ (where N is the pixel super-resolution factor and we set N=4 in this work) to the same dimensionality as the H&E images $x_0 \in R^{H \times W \times 3}$, thereby satisfying the requirements of the subsequent diffusion sampling process:

$$y = f_c(y_0) \tag{1}$$

where $y \in R^{H \times W \times 3}$ is the output of super-resolution header and has a dimension matched with the ground truth image $x_0$.

During the forward diffusion process of Brownian Bridge, the upsampled FTIR image $y$ and the corresponding H&E-stained image $x_0$ were used as the initial state and destination state, respectively, and the intermediate state $x_t$ can be directly computed as follows:

$$x_t = (1 - m_t)x_0 + m_t y + \sqrt{\delta_t}\epsilon_t, \quad \epsilon_t \sim N(0, I) \tag{2}$$

$$m_t = \frac{t}{T}, \quad \delta_t = 2m_t(1 - m_t) \tag{3}$$

where $T$ is the total steps of the diffusion process, and $t$ is the intermediate time step. In this work, $T$ was set to 1000 during both the forward and reverse processes.

Subsequently, the super-resolution header output y is concatenated with the noisy image $x_t$ and fed into a denoising network $x_\theta$, which is trained to estimate $x_0$ from the given $x_t$ and $t$. This denoising network is designed based on a Vision Transformer (ViT) architecture which operates on large image patches consisting of raw pixels. Fig. 2c illustrates the architecture of the pixel diffusion Transformer model, which is consisted of a patchify layer, serval transformer blocks, and a linear decoder layer. The input concatenated image $x_{in} \in R^{H \times W \times C}$ is first reshaped into a sequence of flattened 2D patches $x_p \in R^{N \times (P^2 \times C)}$, where *(H, W)* is the resolution of the original image, C is the number of input channels, P is the resolution of each image patch, and $N = H \times W/P^2$ is the resulting number of patches. In our model, the patch size is set to *P = 16* and the image resolution is 192×192. With an input channel number of C = 6, the patchification process yields a total of $N = (192 \times 192)/16^2 = 144$ image patches. Then after the linear embedding, each patch is transformed into a token of dimension *d*, which is then processed by a series of transformer blocks. Fig. 2c illustrates the detailed design of the transformer block, which consists of a multi-head self-attention layer and an MLP block. The MLP comprises two layers with GELU nonlinear activation functions. Layer normalization (LN) is applied to the image tokens before they are fed into each block. The noise timestep t is embedded into each transformer block through the adaptive layer norm (adaLN). However, a purely Transformer-based architecture may lead to the loss of fine-grained details in the output images, particularly when large patch sizes are used. Therefore, we append a co-trained lightweight U-Net network after the Transformer to recover the fine-grained details of the reconstructed images, as shown in Fig. 2d.

Since we directly predict the clean virtual staining image rather than the noise, the loss function of denoising network $x_\theta$ is defined as:

$$L = \sum_t E_{(x_0, y), \epsilon_t} \left\| x_\theta(x_t, t) - x_0 \right\|^2 \tag{4}$$

In the inference stage, the reverse Brownian Bridge process directly starts from the conditional label-free input by setting $x_T = y = f_c(y_0)$ and aims to predict $x_{t-1}$ from $x_t$:

$$p_\theta(x_{t-1}|x_t, x_0, y) = N\left(x_{t-1}; \mu_\theta(x_t, x_0, t), \tilde{\delta}_t I\right) \tag{5}$$

where the estimated mean value term (learned by the denoising network) is:

$$\mu_\theta(x_t, x_0, t) = a_t x_\theta + b_t x_t + c_t y \tag{6}$$

and the additional variance added during the sampling process is:

$$\tilde{\delta}_t = \frac{\delta_{t|t-1} \cdot \delta_{t-1}}{\delta_t} \tag{7}$$

The expressions of coefficient $a_t$, $b_t$, $c_t$, and $\delta_{t|t-1}$ an be formulated as:

$$a_t = \left(1 - m_{t-1} \frac{\delta_{t|t-1}}{\delta_t}\right) \tag{8}$$

$$b_t = \frac{\delta_{t-1}}{\delta_t} \frac{1 - m_t}{1 - m_{t-1}} \tag{9}$$

$$c_t = \left(m_{t-1} - m_t \frac{1 - m_t}{1 - m_{t-1}} \frac{\delta_{t-1}}{\delta_t}\right) \tag{10}$$

$$\delta_{t|t-1} = \delta_t - \delta_{t-1} \frac{(1 - a_t)^2}{(1 - a_{t-1})^2} \tag{11}$$

Starting from $x_T$, the above steps are repeated to successively estimate $x_{T-1}$, $x_{T-2}$, and so on, until the final clean virtual staining image $x_0$ is obtained. This process can be defined as the following Markov chain:

$$p_\theta(x_{0:T}) = p(x_T) \prod_{t=1}^{T} p_\theta(x_{t-1}|x_t, x_0, y) \tag{12}$$

where each transition probability $p_\theta(\mathbf{x}_{t-1}|\mathbf{x}_t, x_0, y)$ estimates the intermediate state $\mathbf{x}_{t-1}$ from the previous state $\mathbf{x}_t$ and can be formulated as:

$$x_{t-1} = a_t x_\theta + b_t x_t + c_t y + \sqrt{\tilde{\delta}_t}\epsilon, \epsilon \sim N(0, I) \tag{13}$$

Furthermore, to ensure structural consistency and stability of the model outputs, we adopt a mean sampling strategy during inference. Specifically, when the sampling timestep $t > t_e$, the sampling process follows Eq. (12). Once the sampling reaches $t \leq t_e$, the estimation of $x_{t-1}$ depends solely on $x_t$ without injecting additional stochastic noise. In this regime, the sampling process can be expressed as follows:

$$x_{t-1} = a_t x_\theta + b_t x_t + c_t y \tag{14}$$

During the training procedure, both the super-resolution header and the pixel diffusion transformer were trained jointly using a AdamW optimizer with an initial learning rate of 1×10$^{-4}$. The batch size was set at 4 during the training process to balance training speed and GPU memory consumption, and the model required approximately 20 hours to reach convergence. Additionally, during the testing phase, the starting point $t_e$ for mean sampling was set to 10. All network training and testing were performed on a desktop computer with an Intel Xeon W-2235 central processing unit (CPU), 64 GB of random-access memory (RAM), and a Nvidia GeForce GTX 3090 graphics processing unit (GPU). The code for model training was implemented using Python version 3.12.4 and Pytorch version 2.1.0.

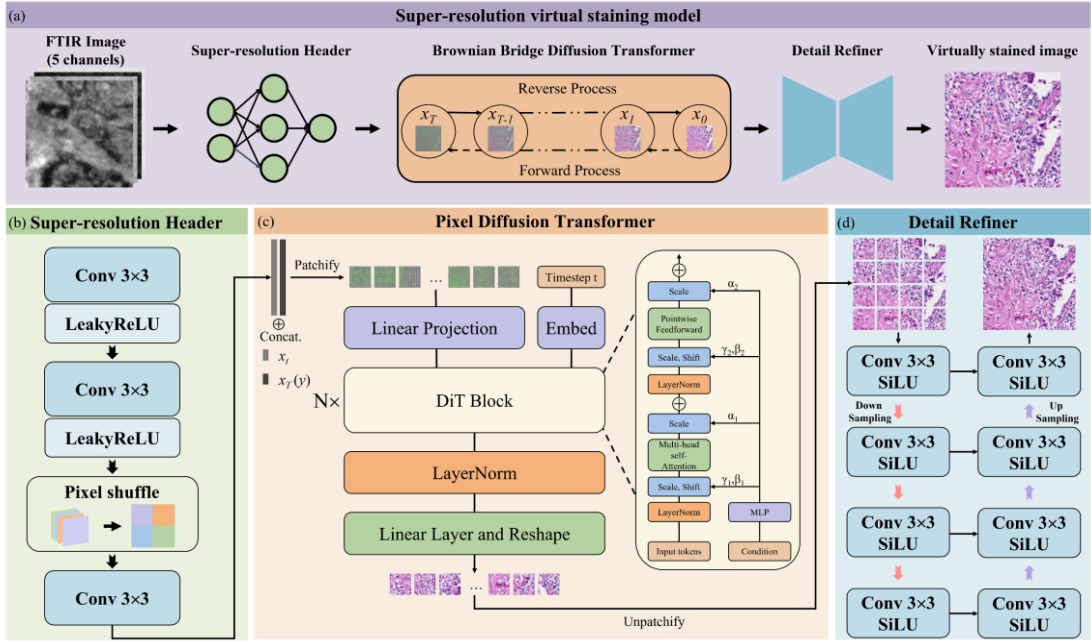

Fig. 2. The network architecture for the DiT-SRVS model. (a) The pipeline of the Diffusion Transformer-based super-resolution virtual staining process; (b)-(d) The detailed architecture of the (b) super-resolution header used for dimension reduction and image upsampling, (c) denoising Transformer backbone, and (d) detail refiner used for detail reconstruction.

## 2.4 Baseline models using cGAN and U-Net based diffusion model

We compared our method with a conventional cGAN-based virtual staining approach and a diffusion-based method employing a U-Net architecture. In the cGAN framework, the generator adopts a five-layer U-Net structure, while the discriminator is implemented as a convolutional neural network-based classifier. The generator is trained using a combination of adversarial loss and pixel-wise structural loss terms, the latter including mean absolute error (MAE) and structural similarity index measure (SSIM). The discriminator is optimized using a least-squares loss function. The learning rate for the cGAN generator is set to $1 \times 10^{-4}$, whereas the learning rate for the cGAN discriminator is set to $1 \times 10^{-5}$. For the diffusion-based virtual staining method with a U-Net backbone, the denoising network is implemented as an attention U-Net and is trained to estimate image noise, following Refs. [32,42]. The learning rate for training the diffusion model is set to $1 \times 10^{-4}$. The AdamW optimizer was employed for network training, and the batch size was set as 4 for all the model training.

## 2.5 Evaluation metrics

To further quantitatively evaluate the performance of H&E virtual staining, we assessed the similarity between the virtually stained images and the corresponding ground-truth H&E-stained images using five widely adopted metrics: Peak Signal-to-Noise Ratio (PSNR), Structural Similarity Index Measure (SSIM), Pearson Correlation Coefficient (PCC), Learned Perceptual Image Patch Similarity (LPIPS), and Fréchet Inception Distance (FID). The core idea of PSNR is to measure distortion by computing the mean squared error (MSE) between the generated image and the ground-truth image, and then quantitatively evaluating it through the ratio of the maximum signal power to the noise power. For a histochemically H&E-stained image X and the corresponding virtually stained image Y of size H×W, the MSE is computed as follows:

$$MSE = \frac{1}{HW} \sum_{i=0}^{H-1} \sum_{j=0}^{W-1} [X(i,j) - Y(i,j)]^2 \qquad (15)$$

and PSNR can be denoted as:
$$PSNR = 20 \cdot log_{10}\left(\frac{MAX_X}{\sqrt{MSE}}\right) \qquad (16)$$
where $MAX_X$ represents the maximum pixel value of the ground truth H&E-stained image.

The SSIM was used to evaluate the perceptual and structural similarity between the virtual and real H&E images. Unlike PSNR, SSIM considers luminance, contrast, and structural information simultaneously, making it more consistent with human visual perception. The SSIM is defined as:
$$SSIM(X,Y) = \frac{(2\mu_x\mu_y + C_1)(2\sigma_{xy} + C_2)}{(\mu_x^2 + \mu_y^2 + C_1)(\sigma_x^2 + \sigma_y^2 + C_2)} \qquad (17)$$
where $\mu_x$ and $\mu_y$ are the mean values of image X and Y. $\sigma_x$ and $\sigma_y$ are the standard deviations of X and Y. $\sigma_{xy}$ is the cross-covariance of X and Y. $C_1$ and $C_2$ are constants introduced to prevent the denominator from being zero.

The PCC was used to quantitatively measure the linear correlation between corresponding pixel intensities of two images. Its value ranges from −1 to 1, where values closer to 1 indicate a strong positive correlation. The PCC is calculated as
$$PCC(X,Y) = \frac{\sum_{i=0}^{H-1}\sum_{j=0}^{W-1}(X(i,j)-\mu_x)(Y(i,j)-\mu_y)}{\sqrt{\sum_{i=0}^{H-1}\sum_{j=0}^{W-1}(X(i,j)-\mu_x)^2}\sqrt{\sum_{i=0}^{H-1}\sum_{j=0}^{W-1}(Y(i,j)-\mu_y)^2}} \qquad (18)$$

LPIPS is a perceptual similarity metric based on deep neural network features extracted from pretrained AlexNet model. Instead of comparing raw pixel values, LPIPS computes the distance between feature representations of two images at multiple semantic levels. A lower LPIPS score corresponds to higher perceptual similarity. The LPIPS score can be expressed as
$$LPIPS(X,Y) = \sum_l \frac{1}{H_l W_l} \sum_{h,w} \left\| w_l \odot (f_l(X)_{h,w} - f_l(Y)_{h,w}) \right\|_2^2 \qquad (19)$$
Where $f_l(\cdot)$ denotes the feature map at layer $l$, $H_l$ and $W_l$ are its spatial dimensions, and $w_l$ represents the learned channel-wise weights.

FID is a metric used to measure the similarity between generated images and real images [53]. It evaluates image quality and diversity by comparing the distributional differences between generated and real images in a feature space. A lower FID score indicates that the distribution of generated images is closer to that of real images, implying that the generated images more closely resemble real images in terms of both visual quality and diversity. The FID was calculated as follows:
$$FID = \left\|\mu_x - \mu_y\right\|^2 + Tr(\Sigma_x + \Sigma_y - 2(\Sigma_x\Sigma_y)^{1/2}) \qquad (20)$$
where $N(\mu_x, \Sigma_x)$ is the multivariate normal distribution estimated from Inception v3 features calculated on real histochemically H&E-stained images and $N(\mu_y, \Sigma_y)$ is the multivariate normal distribution estimated from Inception v3 features calculated on generated virtually stained images. $Tr(\ldots)$ denotes the trace of a matrix.

# 3. Results

## 3.1 Principal Component Analysis of FTIR Spectra

The original FTIR spectral images contain 207 spectral channels, and directly using the raw FTIR data for virtual staining would result in substantial computational overhead. To mitigate this issue, principal component analysis (PCA) was employed for dimension reduction, to reduce the computational cost required for training the virtual staining models. Fig. 3 presents the scree plots obtained from the principal component analysis of the FTIR spectral images collected from six patients, together with the score maps of the first three principal components. The scree plots display the percentage of variance explained by each principal component, enabling an assessment of their relative importance. As shown in the scree plots, the first five principal components account for the majority of the variance in the FTIR data, whereas the variance explained by subsequent components gradually decreases and eventually plateaus, indicating that these later components have limited explanatory power and may contain little useful information. Consequently, we ultimately selected the first five principal components as the training and testing data for the virtual staining models.

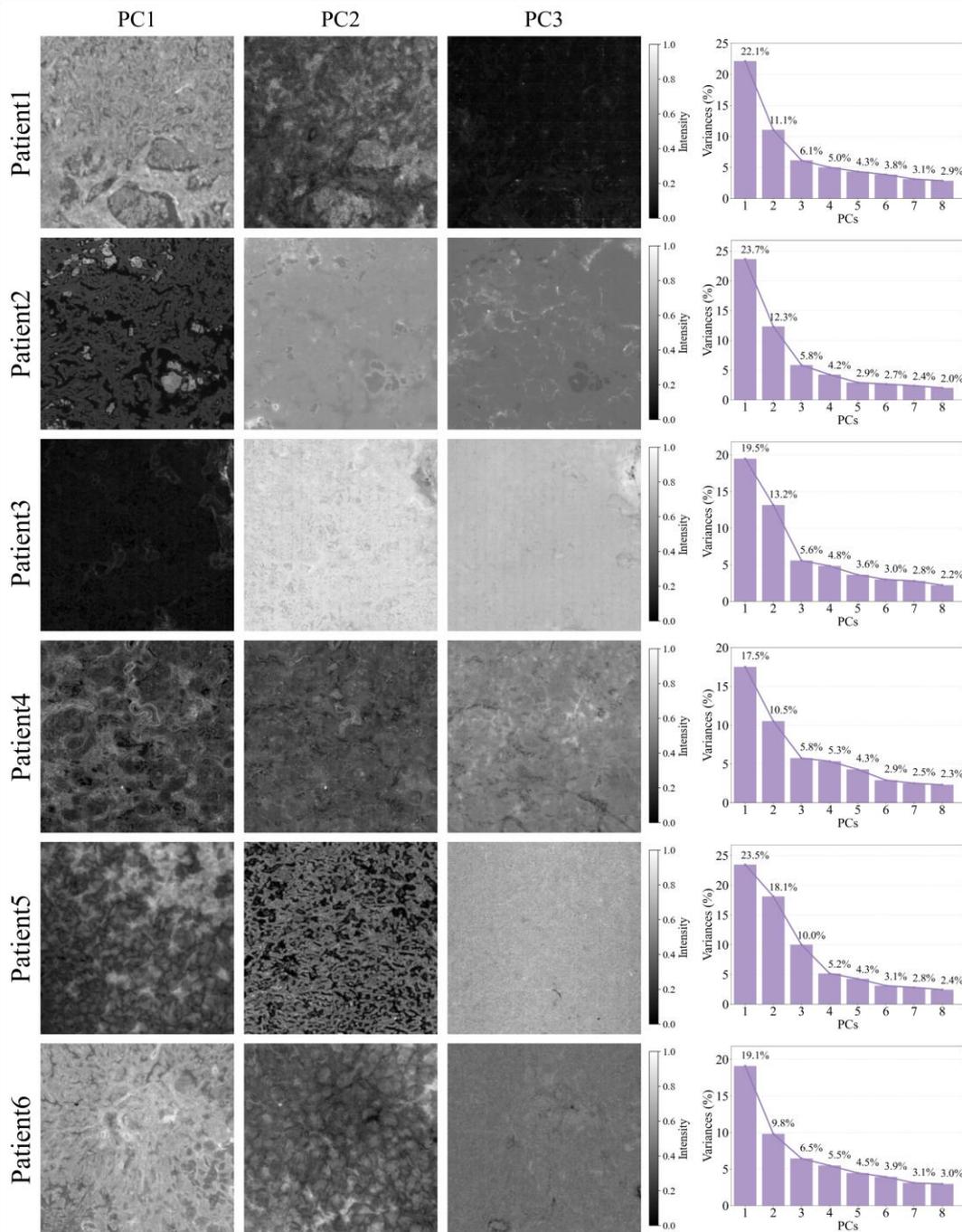

Fig. 3. Principal component analysis results of FTIR spectral images from six patients. The score maps are on the left, where each pixel value represents the projection value of the corresponding spectrum onto that principal component. The scree plots on the right show the percentage of variance explained by each principal component.

## 3.2 Virtual histological staining results of FTIR images of label-free human lung tissues using diffusion Transformer

After the training phase, we evaluated the performance of our DiT-SRVS model on the validation dataset mentioned in section 2.2.3 and compared it with that of the cGAN model and the diffusion U-Net model, as shown in Fig. 4. Fig. 4a-c provides an overview of the FTIR integrated images of the test samples (integrated over the 1660 $cm^{-1} \pm 20$ $cm^{-1}$ band), the corresponding virtual staining results of DiT-SRVS model, and the bright-field microscopic images of the chemically

H&E-stained sections, respectively. Fig. 4d-f shows representative outputs of each of the models. Due to the loss of spatial resolution in FTIR images of unlabeled tissue samples, cGAN-based models are generally unable to accurately reconstruct the stained tissue and cellular structures. In contrast, diffusion-based SRVS models effectively recover these features, which agree well with the corresponding target images. Additionally, we analyzed the color distribution of the virtually stained images generated by each model and compared them with those of the ground-truth H&E-stained images. Fig. 5a presents histograms of color distributions in the YCbCr color space for all test image patches. It can be observed that, among all models, the DiT-SRVS model produces results whose color distributions are closest to the ground truth and exhibit a higher degree of consistency with the histochemically stained target images, demonstrating the superiority of the proposed DiT-SRVS approach.

Fig. 5b-e and Table 1 present the quantitative metrics for each of the models. The comparative analysis demonstrated that diffusion-based virtual staining methods outperform the conventional cGAN model across all evaluated metrics. Moreover, although our DiT-SRVS model achieves slightly lower PSNR, PCC, and LPIPS values than the diffusion U-Net, it is nearly four times faster than the U-Net model in single-image inference test, which can be attributed to its distinctive large-patch input strategy. In addition, we employed paired two-sided *t*-tests to assess whether statistically significant differences in virtual staining performance exist among the three models. The *p*-values shown in Fig. 5b-e highlighted that diffusion-based virtual staining models exhibit statistically significant performance improvements over the traditional cGAN-based approach ($p$-value < 0.05); in contrast, the diffusion U-Net and our DiT-SRVS model demonstrate comparable virtual staining image quality, with no statistically significant differences between them ($p$-value > 0.05).

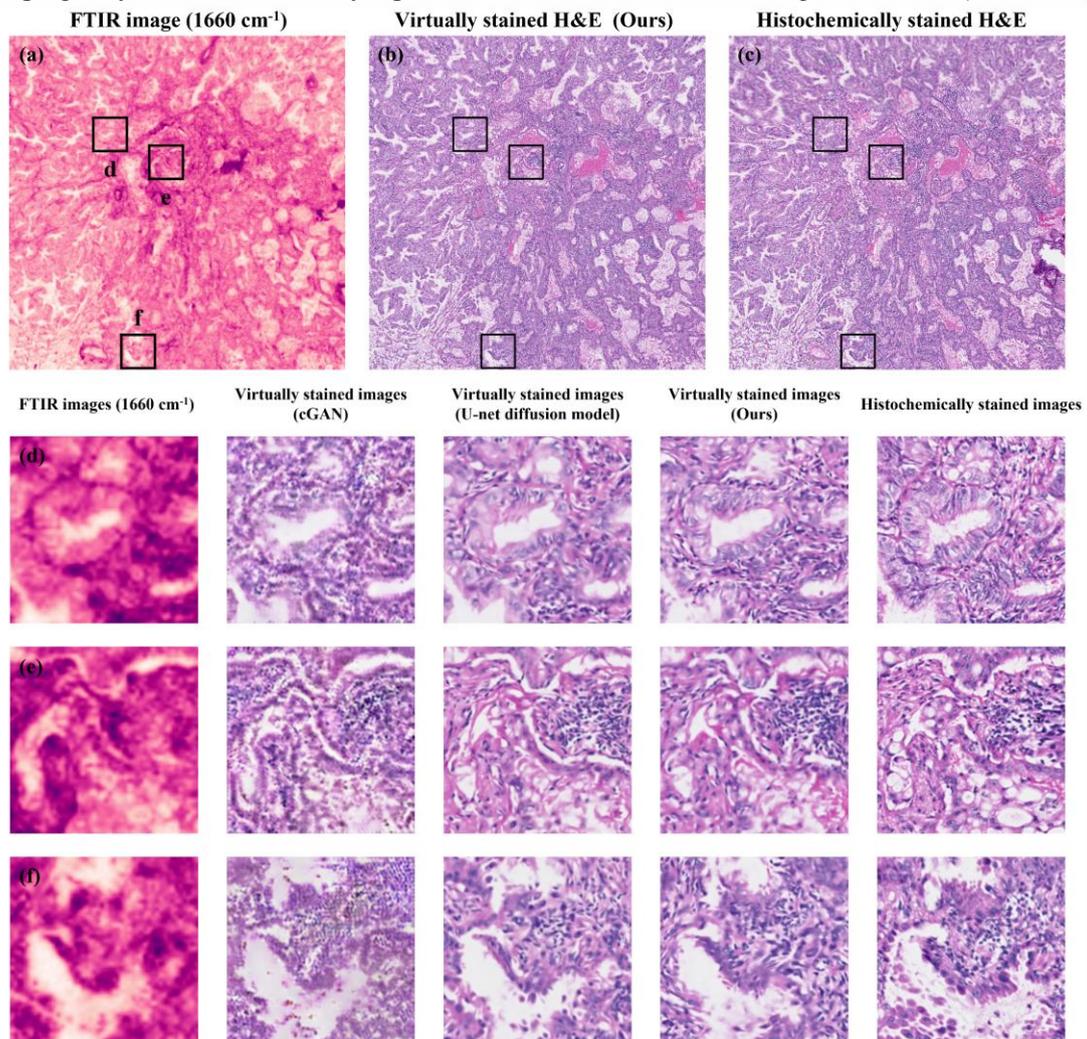

Fig. 4. Visual comparison of super-resolution virtual staining performances of different VS models. (a) FTIR image of the test sample, visualized by integrating over the 1660 cm$^{-1}$ ± 20 cm$^{-1}$ spectral band; (b) Super-resolution virtual staining results of the same FTIR data in (a), which are digitally

generated by our DiT-SRVS model; (c) Histochemical H&E staining results of the tissue section adjacent to (a); (d)-(f) Zoomed-in images of the three exemplary local regions indicated in (a), along with the corresponding super-resolved virtual staining results generated by different models.

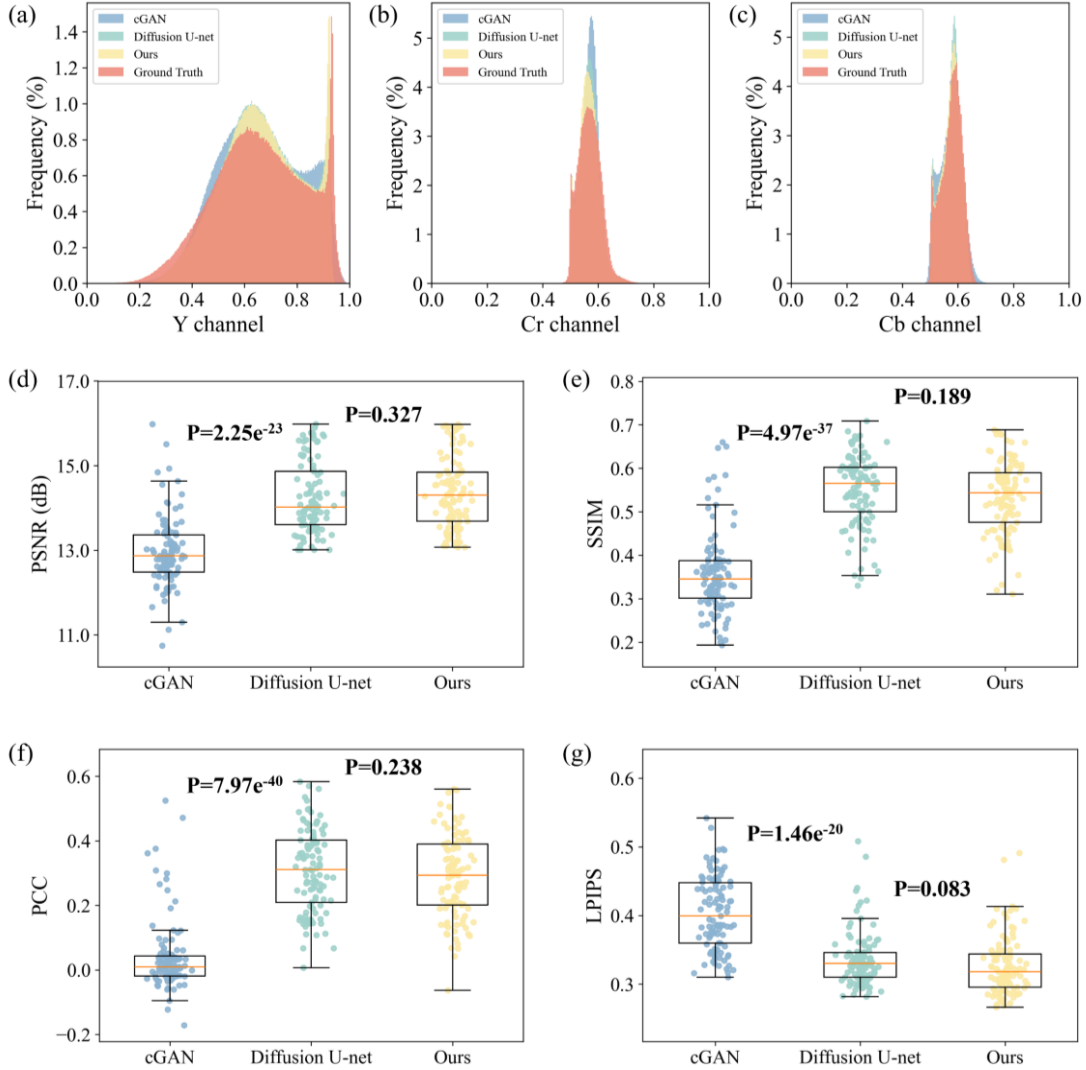

Fig. 5. Quantitative comparison results of super-resolution virtual staining performances of different VS models. (a)-(c) The color histogram comparisons of virtually stained images and their H&E-stained counterparts in (a) Y, (b) Cr, and (c) Cb channels; (d)-(g) The scatter plots present the (d) PSNR, (e) SSIM, (f) PCC, and (g) LPIPS metrics calculated for images generated by different virtual staining methods.

Table 1. Quantitative comparison of different virtual stain methods.

| Metric | cGAN | Diffusion U-Net | Ours |
|---|---|---|---|
| PSNR↑ | 12.95 | 14.24 | **14.36** |
| SSIM↑ | 0.358 | **0.587** | 0.534 |
| PCC↑ | 0.041 | **0.309** | 0.292 |
| LPIPS↓ | 0.404 | 0.337 | **0.326** |
| FID score↓ | 173.66 | 62.98 | **59.53** |
| Latency↓ | **0.29s** | 346.98s | 89.41s |
| Params | 55M | 84M | 236M |

## 3.3 Ablation analysis on the detail refiner

The Detail Refiner (DR) is a submodule composed of a lightweight U-Net that helps recover the fine details in the images generated by the vision Transformer, thereby further enhancing the quality of virtual staining. To further shed light on this, we analyzed the staining performance of the DiT-SRVS model with and without the DR module on the test dataset, and the comparative results are presented in Fig. 6 and Table 2. Although the DR module accounts for less than 2% of the total model parameters, incorporating it into the DiT-SRVS model leads to improvements across all image evaluation metrics compared with the model without the DR module, including a 30% improvement in the FID score. Moreover, the results of the two-sided *t*-tests indicate that the model with the DR module achieves statistically significant performance gains over its counterpart without the module in terms of SSIM and LPIPS metrics (*p*-value < 0.05). These results confirm that the DR module enables a statistically significant improvement in the virtual staining fidelity of the DiT-SRVS model, highlighting its strong practical utility for virtual staining applications in unlabeled tissue imaging.

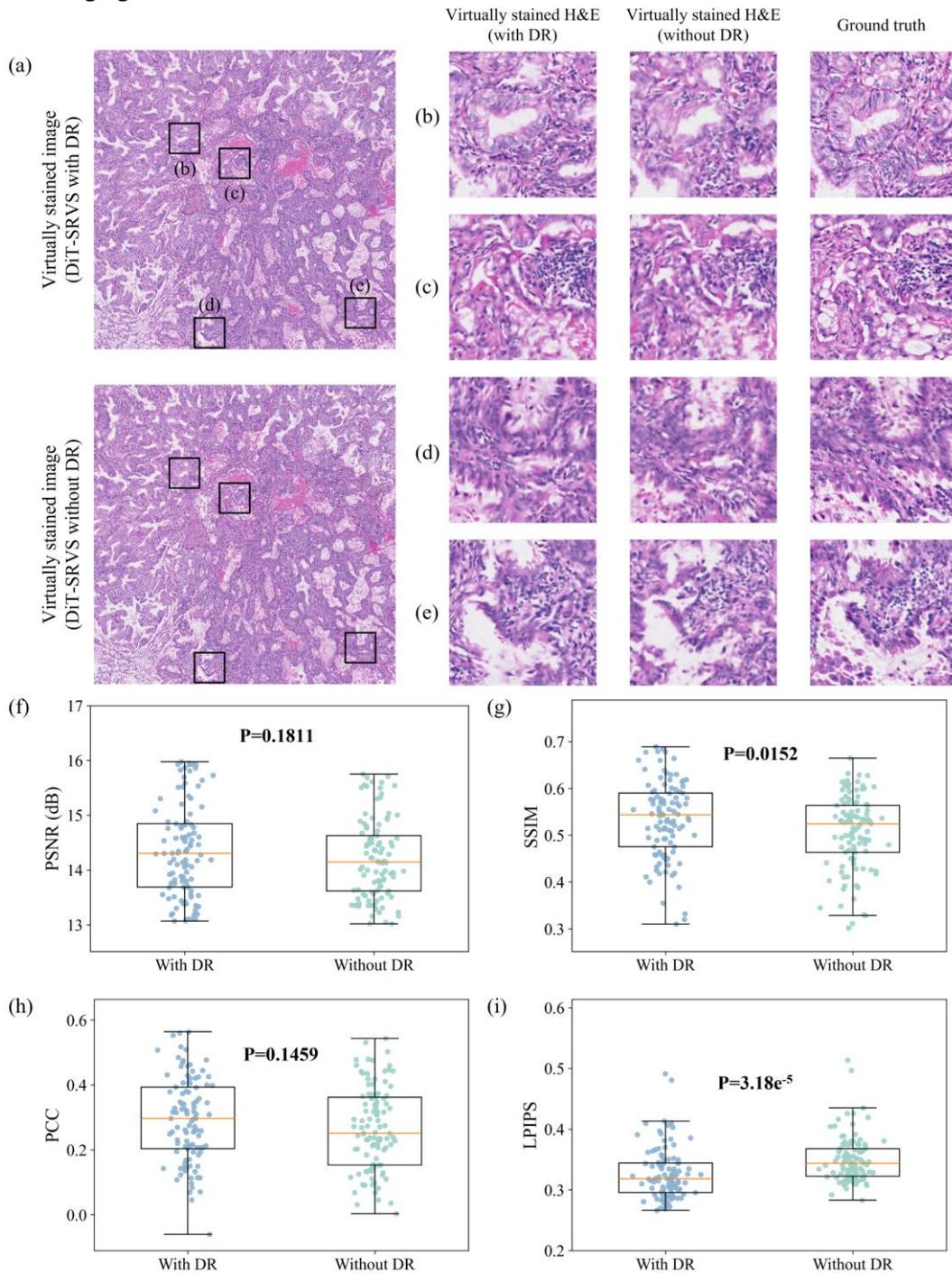

Fig. 6. Comparison of VS performance for DiT-SRVS models with and without the DR module. (a)

Visual comparisons of virtually stained images generated from DiT-SRVS models with and without the DR module; (b)-(e) Zoomed-in images of the four exemplary local regions indicated in (a), along with the corresponding ground truth; (f)-(i) The scatter plots present the (f) PSNR, (g) SSIM, (h) PCC, and (i) LPIPS metrics calculated for images generated by DiT-SRVS models with and without the DR module.

Table 2. Quantitative comparison of DiT-SRVS model with or without DR module.

| Metric | DiT-SRVS (with Detail Refiner) | DiT-SRVS (without Detail Refiner) |
|---|---|---|
| PSNR↑ | **14.36** | 14.21 |
| SSIM↑ | **0.534** | 0.506 |
| PCC↑ | **0.292** | 0.281 |
| LPIPS↓ | **0.326** | 0.352 |
| FID score↓ | **59.53** | 85.16 |
| Latency↓ | 89.41s | **72.59s** |
| Params | 236M | 232M |

# 4. Conclusion

In this work, we proposed a diffusion Transformer–based super-resolution virtual staining (DiT-SRVS) model for converting low-resolution FTIR spectral images of unstained tissue into corresponding high-resolution H&E-stained images. The model adopts a Transformer backbone that takes large image patches as input and learns the domain translation directly in pixel space through a Brownian bridge diffusion process. In addition, the overall framework incorporates a super-resolution head and a detail refiner: the former upsamples the input FTIR images and performs dimensional transformation to match the spatial resolution of the H&E-stained images, while the latter further refines fine-grained details in the Transformer outputs, thereby enhancing the quality of the generated virtual staining images.

By effectively combining the strengths of diffusion models and large-patch Transformers, our method demonstrates strong performance on experiments conducted with label-free human lung tissue. Both qualitative visual inspection and quantitative evaluation show that the proposed DiT-SRVS model significantly outperforms conventional cGAN-based methods and achieves performance comparable to current state-of-the-art approaches. Moreover, the use of large-patch inputs substantially reduces computational overhead and inference latency; in single-image inference tests, DiT-SRVS is approximately four times faster than traditional diffusion models based on U-Net architectures.

Overall, our approach provides a reliable and efficient framework for rapidly generating directly corresponding H&E-stained images from FTIR data, thereby eliminating the need for subsequent complex histochemical staining and image registration procedures and greatly accelerating the turnaround of clinical infrared metabolomics studies. In the future, we will further enhance the staining performance and generalization capability of DiT-SRVS by training and evaluating the model on broader and more diverse datasets.

Acknowledgements
This research was sponsored by the National Natural Science Foundation of China (62375170 and 62535019), the Shanghai Jiao Tong University (YG2024QNA51), and the Science and Technology Commission of Shanghai Municipality (20DZ2220400). We thank Dr. Haitang Yang from Shanghai Chest Hospital of Shanghai Jiao Tong University for providing lung tissue samples.